\documentclass[aps,prb,twocolumn,preprintnumbers,amsmath,amssymb,citeautoscript]{revtex4-1}
\usepackage{graphicx}
\usepackage{wrapfig}
\usepackage{amsmath}
\usepackage{color, soul}
\usepackage{hyperref}
\usepackage{enumerate}

\def\circled#1{\textcircled{\mbox{\tiny #1}}}
\def\ul#1{\underline{#1}}
\def\sm#1{{\mbox{\tiny #1}}}
\def\ve{{\bf e}}
\def\vj{{\bf j}}
\def\vp{{\bf p}}
\def\vq{{\bf q}}
\def\vr{{\bf r}}
\def\vu{{\bf u}}
\def\vv{{\bf v}}
\def\vx{{\bf x}}
\def\vy{{\bf y}}
\def\vA{{\bf A}}
\def\vR{{\bf R}}
\newcommand{\SF}{F_\sm{sf}}
\newcommand{\hk}{{\hat{k}}}
\newcommand{\hp}{{\hat{p}}}
\newcommand{\Tc}{T_{\mathrm{c}}}
\newcommand{\vf}{v_\sm{F}}
\newcommand{\vvf}{\vv_\sm{F}}
\newcommand{\vpf}{\vp_\sm{F}}
\newcommand{\Nf}{N_{\mathrm{F}}} 
\newcommand{\Del}{{\mbox{\footnotesize $\Delta$}}}
\newcommand{\grad}{\mbox{\boldmath$\nabla$}}

\newcommand{\dive}{\mbox{\boldmath$\nabla$}\cdot}

\newcommand{\vare}{\varepsilon}
\newcommand{\sgn}{\mbox{sgn}}
\newcommand{\be}{\begin{equation}}
\newcommand{\ee}{\end{equation}}
\newcommand{\He}{$^3$He}

\graphicspath{{./}{./pict/}}
\begin{document}
\title{Phase Crystals}

\author{P. Holmvall}
 \affiliation{Department of Microtechnology and Nanoscience - MC2, Chalmers University of Technology, SE-41296 G{\"o}teborg, Sweden}
\author{M. Fogelstr{\"o}m}
 \affiliation{Department of Microtechnology and Nanoscience - MC2, Chalmers University of Technology, SE-41296 G{\"o}teborg, Sweden}
\author{T. L{\"o}fwander}
 \affiliation{Department of Microtechnology and Nanoscience - MC2, Chalmers University of Technology, SE-41296 G{\"o}teborg, Sweden}
\author{A. B. Vorontsov}
\email[]{anton.vorontsov@montana.edu}
  \affiliation{Department of Physics, Montana State University, Montana 59717, USA}

\date{\today}

\begin{abstract}

Superconductivity owes its properties to the phase of the electron pair condensate that breaks the $U(1)$ symmetry.
In the most traditional ground state, the phase is uniform and rigid. 
The normal state can be unstable towards special inhomogeneous superconducting states: 
the Abrikosov vortex state, and the Fulde-Ferrell-Larkin-Ovchinnikov state. 
Here we show that the phase-uniform superconducting state can go into a fundamentally different and  
more ordered non-uniform ground state, that we denote as a phase crystal. 
The new state breaks translational invariance through formation of a spatially periodic modulation of the phase, 
manifested by unusual superflow patterns and circulating currents, that also break time-reversal symmetry.
%
We list the general conditions needed for realization of phase crystals. 
Using microscopic theory we then derive an analytic expression for the superfluid density tensor for the case of a non-uniform 
environment in a semi-infinite superconductor. We demonstrate how the surface quasiparticle
states enter the superfluid density and identify phase crystallization as the main player in several 
previous numerical observations in unconventional superconductors, and predict existence of a similar phenomenon in
superconductor-ferromagnetic structures. 
This analytic approach provides a new unifying aspect for the exploration of boundary-induced quasiparticles and
collective excitations in superconductors. 
More generally, we trace the origin of phase crystallization to 
non-local properties 
of the gradient energy, 
which implies existence of similar 
pattern-forming instabilities in many 
other contexts.


\end{abstract}

\maketitle

\section{Introduction}
%
The defining characteristic of superfluidity and superconductivity 
is spontaneous symmetry breaking of the global $U(1)$ phase $\chi$, associated with the order parameter
$\Delta = |\Delta|\exp(i\chi)$.
The phase, and its spatial variations, give rise to phenomena of importance for technological applications, such as 
type~II superconductivity where Abrikosov vortices are formed in an external magnetic field, and in Josephson junctions \cite{Tinkham}.
Within the BCS paradigm \cite{BCS}, a uniform fixed value of the phase is directly tied to the finite amplitude $|\Delta|$ of 
the macroscopic Cooper-pair wavefunction.
If the phase is non-uniform, by Galilean invariance it results in superflow with superfluid velocity and momentum 
$m \vv_s = \vp_s(\vR) = (\hbar/2) \grad \chi(\vR)$, where $m$ is the electron mass and $\hbar$ is the reduced Planck constant.
Such phase variations and the associated condensate currents cost gradient energy
\be
\SF = \frac12 \int d\vR \; k|\Delta|^2 \, | \grad \chi (\vR)|^2 \,,
\ee
where the gradient energy coefficient $k>0$ should be computed from microscopic theory. 
A physical picture emerges where the phase is rigid, 
coherent over macroscopic distances, 
and the superconducting state is stable.
Thus, it would be surprising if there existed a more ordered state with a softer phase and 
spontaneous superflow with energy gain $\SF<0$.
%

\begin{figure*}[t]
\hfill
\includegraphics[width=0.65\textwidth]{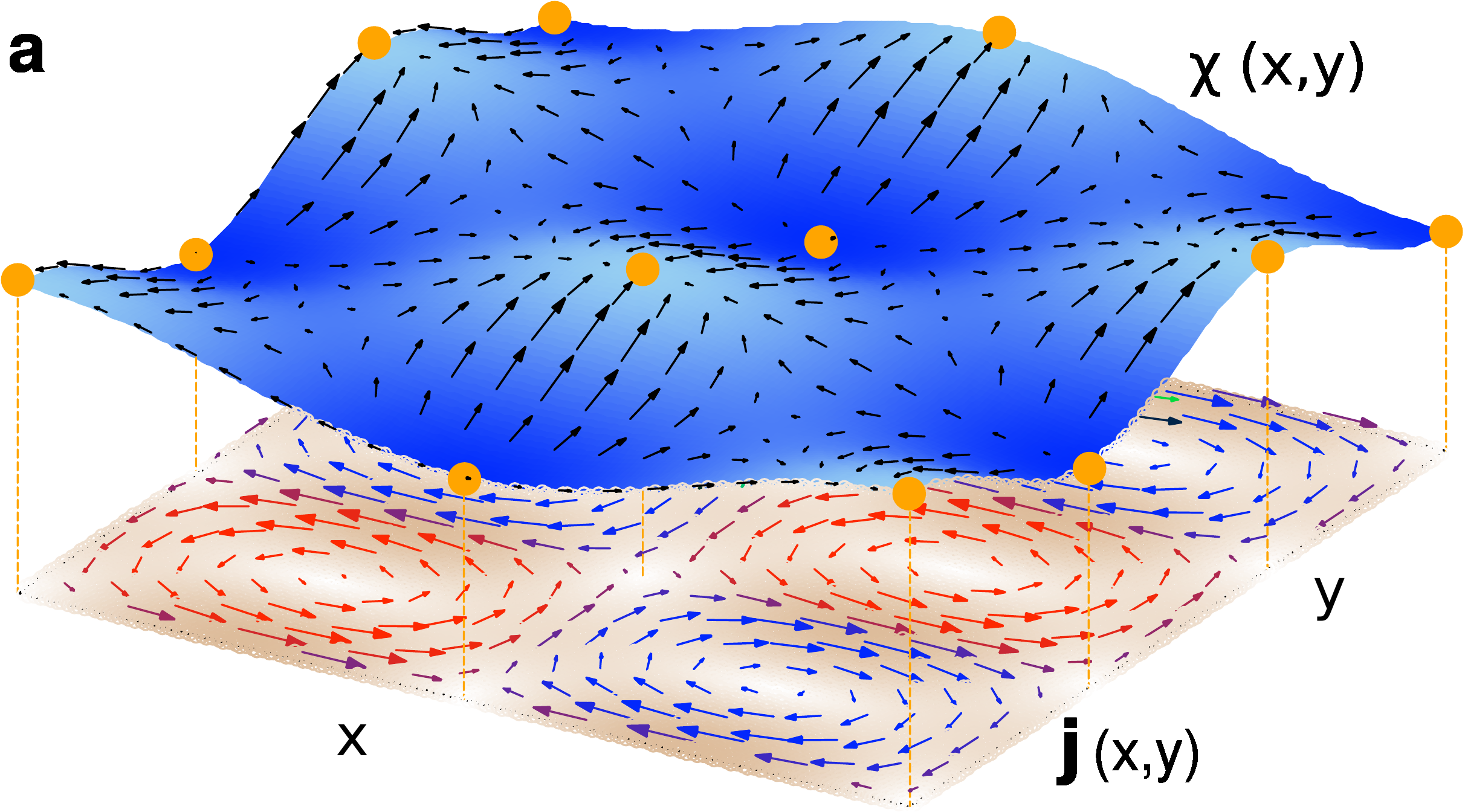}
\hfill
\raisebox{6mm}{\includegraphics[width=0.28\textwidth]{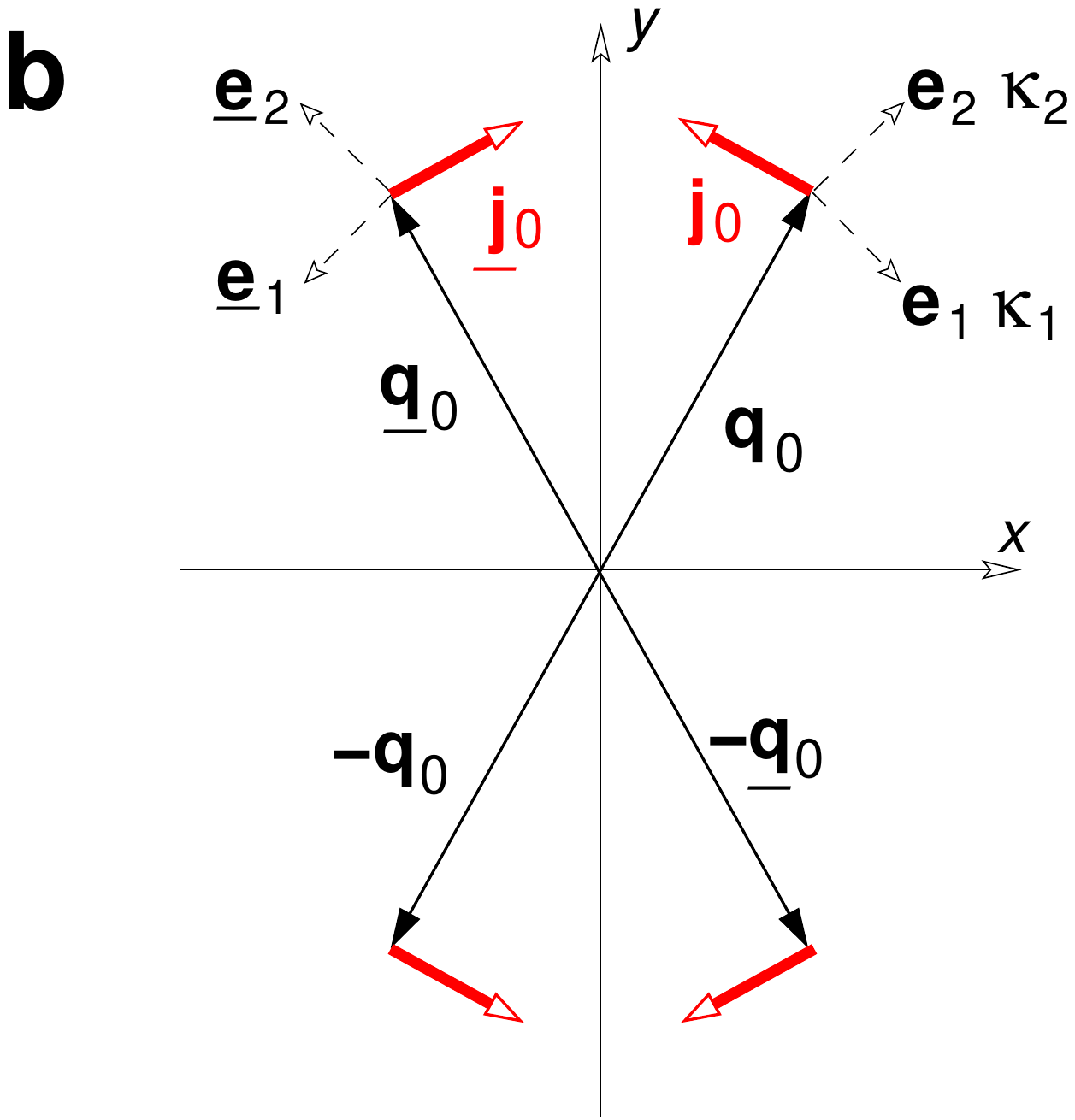}}
\caption{{\bf a}, The phase crystal has a periodic modulation of the superconducting phase $\chi(\vR)$ 
and a superflow $\vp_s(\vR)$ 
that forms a special vector field with a lattice of sources and sinks (filled circles),
while the particle-conserving current $\vj(\vR)$ forms a checkerboard pattern with opposite circulation flow.
{\bf b}, This phase modulation is a result of four degenerate instability vectors $\{\pm\vq_0, \pm \underline{\vq}_0\}$ 
with non-zero currents orthogonal to them, see Eq.~(\ref{bulkJ}). 
}
\label{fig:phasecrystal}
\end{figure*}

Here, we propose that under certain conditions 
there exists a low-temperature superconducting state where 
the rigid phase acquires structure 
by breaking translational invariance. 
In this state, that we denote a phase crystalline state, a periodic pattern with 
wavevector $\vq$ is formed
\be
\chi(\vR) = C_\vq \, A_\vq(\vR_\perp) \, \cos (\vq\cdot \vR),
\label{chi_phasecrystal}
\ee
where $A_\vq(\vR_\perp)$ is a function of coordinates orthogonal to $\vq$.
The additional order parameter in the phase crystal is the finite Fourier amplitude $C_\vq$. 
The superconducting ground state with spatially oscillating phase also breaks time-reversal symmetry and sustains a non-trivial periodic superflow
pattern and circulating currents $\vj(\vR)$, as illustrated in Fig.~\ref{fig:phasecrystal}{\bf a}.
Similar current patterns have been found in numerical work on mesoscopic grains of $d$-wave superconductors \cite{Hakansson:2014uf},
and the unusual superflow field $\vp_s(\vR)$ was recently analyzed\cite{Holmvall:2018fl}.
Here we establish that the physical origin of this 
surface state is phase crystallization.

Breaking of continuous translational symmetry 
is particularly striking. Its reduction to discrete translations gives
a multitude of crystals \cite{Powell} and ultimately quasicrystals where translational symmetry is absent
\cite{Senechal,*symm_crystal,Kats:1993,Martin2016}.
Crystal analogues in the time dimension \cite{time_crystal,Yao2017} have been recently observed \cite{Zhang:2017uw,Choi:2017wn}.
Emergent multi-particle crystalline structures are predicted to appear in frustrated magnetic materials,\cite{Kamiya:2016} 
and have been engineered in ultracold atoms interacting with light.\cite{Ostermann:2016}
Superconducting states with periodically modulated amplitude $\Delta(\vR) \propto \Delta_\vq \cos(\vq\cdot\vR)$ 
were first proposed to exist in ferromagnetic metals
\cite{Larkin:1964uu,*Larkin:1965wj}, 
and are currently investigated in a variety of systems ranging from 
cold Fermi-gases with spin imbalance \cite{Kinnunen2018,Dutta2017} 
to color superconductivity \cite{FFLO_QCD}.

Several features make the phase crystal a distinctly different ground state from other non-uniform superconducting states. 
The amplitude-modulated state and its single-mode \cite{FuldeFerrell} counterpart 
$\Delta(\vR) \propto \Delta_\vq e^{i\vq\cdot\vR}$,
are both \emph{amplitude} instabilities of the normal metal occurring at finite $\vq$, 
and they do not carry currents.
The phase crystal, on the other hand,  
is associated with a modification of 
the symmetry variable $\chi$ describing the 
degeneracy manifold of the superconducting state, 
and can occur even when the order parameter amplitude  $|\Delta|$ is large, i.e. deep inside the superconducting state
far from the normal to superconductor transition; 
the phase crystal does maintain non-trivial particle currents. 
%
Moreover, it is also different from the textures appearing in systems with multi-component  
order parameters and a more complex degeneracy space, such as \He\ and liquid crystals\cite{vol90,Chaikin,deGennes}.
In those systems the long-wavelength textures are a result of a competition 
between condensation and gradient terms involving different combinations of the order parameter components. 
%
The phase crystal is a result of a highly non-local superfluid response 
when sample surfaces, geometry, or other external influences, 
impose a certain structure on the superfluid kernel itself.
The patterns are formed on the much shorter coherence length scale
$\xi_0 = \hbar \vf/ 2\pi k_{\mathrm{B}} \Tc$, where $\vf$ is the Fermi velocity, $\Tc$ is
the superconducting transition temperature and $k_{\mathrm{B}}$ is the Boltzmann constant 
($\hbar=k_{\mathrm{B}}=1$ in the following).  
To describe 
this physics we ignore the amplitude gradient terms in the free energy and generalize 
the kinetic superflow energy in the limit of small $\vp_s$ as
\begin{align}\begin{split}
\SF[\grad\chi] = \frac12 \iint d\vR d\vR'  \; \nabla_i \chi(\vR) K_{ij}(\vR,\vR') \nabla_j \chi(\vR') \,,
\label{SuperflowFE}
\end{split}\end{align}
where we introduce a non-local superfluid density kernel $K_{ij}(\vR,\vR') = K_{ji}(\vR',\vR)$. 
Summation over repeating 
spatial indices is assumed.
Higher order gradient terms in $\SF$ would determine the magnitude of spontaneous currents at temperatures below the transition temperature.
Here we neglect those and focus on the instability analysis.
\footnote{
	We also drop corrections to the superflow 
	due to the vector potential $\vA(\vR)$ of the self-induced field $\grad\chi \to  \grad \chi - \frac{2\pi}{\Phi_0} \vA$. 
	These corrections result in energy terms that are smaller than the phase-gradient terms by factor 
	$(\xi_0/\lambda)^2$, which is small in type-II superconductors. See e.g. Refs. \onlinecite{Holmvall:2018fl,Barash:2000vt} 
}
The energy change 
due to a small Galilean boost $\vu$, 
$\SF[\vv_s -\vu] = \SF[\vv_s] - m \vj \cdot \vu$,  
defines the particle current 
\begin{align}\begin{split}
j_i(\vR) = \frac{\delta \SF[ \vv_s] }{\delta p_{s,i}(\vR)} = 
\int d\vR'  K_{ij}(\vR,\vR') \nabla_j \chi(\vR') \,.
\label{SuperflowJ}
\end{split}\end{align}
The physical $\chi$ and $\vj$ are obtained by variational minimization of the free energy with respect to the phase. 
It gives the continuity equation, 
$ - { \delta \SF[\grad \chi]}/{\delta \chi(\vR)} = 
\dive \vj(\vR) = 0$. 

\section{phase instability in the bulk} 
%
By using the non-local Ginzburg-Landau expression in Eq.~(\ref{SuperflowFE}) one can
specify the general criteria 
when a non-trivial pattern of currents can emerge from the state with homogeneous phase $\chi_0=0$. 
In a translationally-invariant infinite system 
the superfluid free energy with kernel $\hat K(\vR-\vR')$ has the following form in Fourier space
\begin{align}\begin{split}
\SF  =  \frac12 \int \frac{d^2q }{(2\pi)^2} \;  \chi(-\vq) \, \left[ \vq^T \hat K(\vq) \, \vq \right] \; \chi(\vq)  \,.
\end{split}\label{FEbq}\end{align}
For the two-dimensional case,
the kernel is a two-by-two Hermitian matrix $\hat K(\vq) = \hat K^\dag(\vq)$ with 
real eigenvalues $\kappa_{1,2}$ and corresponding eigenvectors $\ve_{1,2}$. Their values depend on temperature and $\vq$. 
The instability at a particular wavevector $\vq_0$ can happen when 
$\vq_0^T \hat K(\vq_0) \vq_0 = \kappa_1 [\ve_{1} \cdot \vq_0]^2+  \kappa_2 [\ve_{2} \cdot \vq_0]^2 = 0$.
This equality can be satisfied if the eigenvalues have opposite signs and are tunable by temperature, 
or more generally by some other parameter. 
To linear order in $\chi(\vq)$, the Fourier component of the current is
$\vj = \vj_0 \, i \,  \chi(\vq_0)$, 
where $i=\sqrt{-1}$ and $\vj_0 = \hat K(\vq_0) \vq_0 = \ve_{1} \kappa_1 [\ve_{1} \cdot \vq_0] +  \ve_{2} \kappa_2 [\ve_{2} \cdot \vq_0]$.
For a non-zero current to appear at the $\vq_0 \ne 0 $ transition, it must also satisfy 
the conservation law $\dive \vj \propto \vq_0 \cdot \vj = 0$. 
This implies an orthogonality constraint
$\vq_0 \perp \vj_0$,
which is possible to fulfill if the eigenvectors $\ve_{1,2}$ are not collinear with $\vq_0$, 
see Fig.~\ref{fig:phasecrystal}{\bf b}.
In this case we can write $\vj_0 = \hat\vx j_{0x} + \hat\vy j_{0y}$ with $j_{0x} / j_{0y} = -q_{0y}/q_{0x}$.
Since the phase $\chi(\vR)$ is real, the same conditions must be satisfied for $-\vq_0$, which 
requires inversion symmetry. With two instability vectors $\vq_0$ and $-\vq_0$ we get 
an emerging phase $\chi(\vR) = C \cos(\vq_0 \cdot \vR)$ with stripes of current 
$\vj(\vR) = C \vj_0 \sin(\vq_0 \cdot \vR)$ running perpendicular to $\vq_0$. 
Additional symmetries allow for other instability vectors.
For example, reflection symmetry $x \to -x$ guarantees another pair of instability vectors, 
$\ul\vq_0$ and $-\ul\vq_0$, with $\ul{q}_{0x} = -q_{0x}$. 
Diagonalization of the kernel at $\ul\vq_0$ gives the same eigenvalues $\kappa_{1,2}$ as
those at $\vq_0$, while the eigenvectors $\ul\ve_{1,2}$ are obtained from $\ve_{1,2}$ by flipping the $x$-components, 
and the current amplitude is 
$\ul\vj_0 = \ul\ve_{1} \kappa_1 [\ul\ve_{1} \cdot \ul\vq_0] +  \ul\ve_{2} \kappa_2 [\ul\ve_{2} \cdot \ul\vq_0]$. 
In the four-harmonics state the phase and current are given by 
\begin{align}\begin{split}
\chi(\vR) & =  \cos(\vq_0 \cdot \vR) + \cos(\ul\vq_0 \cdot \vR)   \propto  \cos(q_{0x} x) \cos(q_{0y} y),
\\
\vj(\vR) & = \vj_0 \sin(\vq_0 \cdot \vR) + \ul\vj_0 \sin(\ul\vq_0 \cdot \vR) 
\\
&\propto \begin{pmatrix} 
j_{x0} \sin(q_{0x} x) \cos(q_{0y} y) \\
j_{0y} \cos(q_{0x} x) \sin(q_{0y} y)
\end{pmatrix},
\end{split}\label{bulkJ}\end{align}
as plotted in Fig.\ref{fig:phasecrystal}{\bf a}.
Higher order terms ${\cal O}[(\grad \chi)^4]$ must be included to determine the energetics between two- and
four-harmonics states.  
One notices that the loop currents in the phase crystal appear without phase winding and are not associated with topological defects. 
We conclude that 
realization of spontaneous periodic loop-currents requires a superfluid density tensor with 
\begin{enumerate}[(i)] 
\itemsep-1mm
\item spatial anisotropy,
\item positive and negative eigenvalues that can be tuned by some parameter,
\item eigenvectors $\ve_{1,2} \nparallel \vq_0$. 
\end{enumerate}
Conditions (i) and (ii) can be satisfied simultaneously for example in an anisotropic-gap superconductor with an applied Zeeman field.
Condition (iii) requires a mismatch between the symmetry of the Fermi surface and the quasiparticle excitations in
momentum space, and the symmetry of the current response tensor.
To satisfy this last geometric condition, one would generally require 
a system with as lower spatial symmetry as possible. To formalize the analysis we can 
write a general Ginzburg-Landau expansion of the tensor $\hat K(\vq)$ 
in the superconducting state with orthorhombic symmetry $C_{2v}$.  
This symmetry is also required by condition (i) to have two eigenvectors of the kernel of different sign. 
The general form of the tensor is 
\begin{align}\begin{split}
K_{ij}(q_x,q_y) & = K_{ij}^{(0)} + K_{ijlm}^{(2)} q_l q_m + \dots =  
\\
& = \begin{pmatrix}
a_0 + a_2 q_x^2 + c_2 q_y^2   &    2c_2 q_x q_y  
\\
    2 c_2 q_x q_y   &      b_0 + b_2 q_y^2 + c_2 q_x^2
\end{pmatrix}
\label{Ktensor}
\end{split}\end{align}
where finite components are $a_0 = K_{xx}^{(0)} \ne K_{yy}^{(0)}=b_0$, 
$K_{xxxx}^{(2)} = a_2$, 
$K_{yyyy}^{(2)} = b_2$, 
$K_{xxyy}^{(2)} = c_2$, and all permutation of indices allowed.
The configuration space of these five coefficients is large enough to  allow for a set of instability wavevector $(q_x,q_y)$ 
that do not lie along the high symmetry directions, and thus do not coincide with direction of the current $(j_x, j_y)$. 
Such configuration would not be possible in a state with square symmetry that has only three independent coefficients
$a_0=b_0$, $a_2=b_2$ and $c_2$. 
The superfluid tensor will possess the $C_{2v}$ symmetry in orthorhombic crystals, 
in nematically ordered systems, or in superconducting states with gap structure different along two principal axes, 
such as polar or planar states. 
The complete analysis of a crystallization transition with a short-wavelength modulations is quite complex, 
and has to include higher order $\vq$-terms. 
We leave this for future studies. 
We note that in typical weak crystallization theories the instability
vectors are only given at phenomenological level.\cite{Kats:1993,Martin2016} 
In the following we write down the microscopic theory for $\hat K$  near pairbreaking  surfaces and show 
how all these conditions are naturally satisfied 
and why a preferred ordering vector emerges.


\begin{figure}[t]
\centering\includegraphics[width=0.95\linewidth]{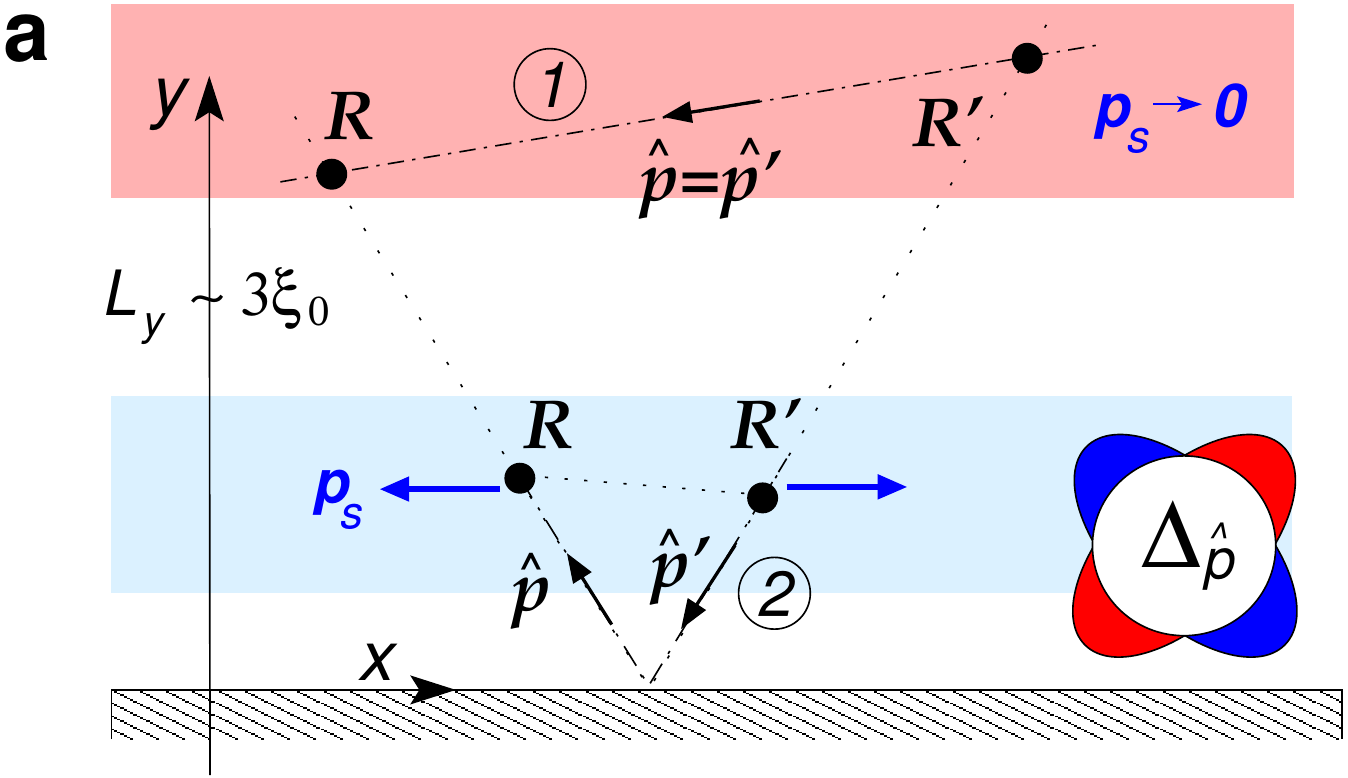}
\centering\includegraphics[width=0.95\linewidth]{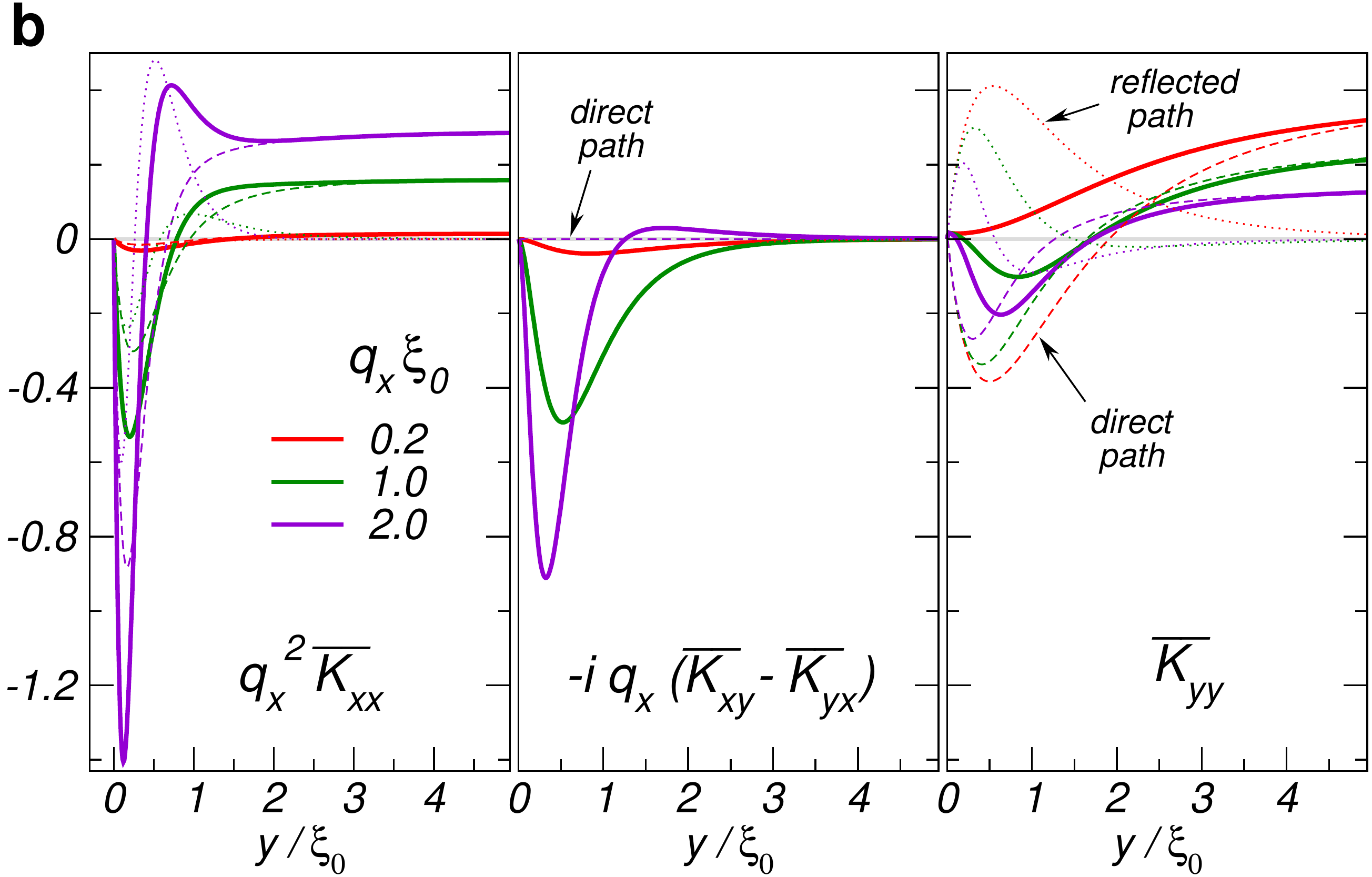}
\caption{
	{\bf a}, Microscopic model of the superfluid density tensor near a pairbreaking surface of a $d_{xy}$ superconductor. 
	{\bf b}, The averaged `local' components, Eq.~(\ref{Kavrg}), 
	as a function of distance to the surface $y$ and the modulation vector $q_x$.
	%
	%
	The thinner dashed lines show direct path's contribution, dotted - reflected path.
	The superfluid density far from the surface is determined by correlations between two points, 
	$\vR$ and $\vR'$, through the direct path. 
	This leads to positive superflow energy from diagonal components, 
	favoring a uniform phase $\vp_s \propto \grad \chi_0 = 0$. 
	Near the surface the superflow energy is lowered by negative contributions of $K_{xx}$ and $K_{yy}$ coming from
	Andreev bound states,
	favoring the non-uniform phase crystal $\grad \chi \ne 0$.
}
\label{fig:S}
\end{figure}

\section{Surface Phase Crystal} 
%
Using microscopic quasiclassical theory, we derive the general expression for the superfluid
density kernel. 
The technical details of the calculation are moved to Appendix \ref{appA}.
We apply it first to the $d$-wave case and consider the $s$-wave case at the end of this section.
The $d$-wave superconductor has an order parameter 
$\Delta(\vR,\vpf) = \Delta_0(\vR) \, [2\hp_x\hp_y] \equiv \Delta_\hp$, 
oriented as shown in Fig.~\ref{fig:S}{\bf a}. 
The $\hat p = \vpf/|\vpf|$ is the unit vector pointing in the direction of momentum $\vpf$ on the Fermi surface.
The kernel between two points $\vR$ and $\vR'$ in a semi-infinite system has two contributions,  
$\hat K(\vR,\vR') = \hat K^{\circled{1}}(\vR,\vR') + \hat K^{\circled{2}}(\vR,\vR')$, that correspond to propagation of
quasiparticles along the direct path or with a reflection at the surface. 
We set a uniform amplitude $\Delta_0(\vR)=\Delta_0$, 
which allows for analytic expressions, Appendix \ref{appB}. 
This assumption also demonstrates that the phase crystal is not caused by the suppression of the order parameter per se, 
but rather by the contribution from the symmetry-related surface Andreev bound states. 
The coordinate along a quasiparticle trajectory is denoted by $s$, with $s=0$ at the reflection point. 
The kernel components are calculated in Appendix \ref{appC}, and for the direct path ($\hp' = \hp$) they are
\begin{align}\begin{split}
K^{\circled{1}}_{ij}(\vR,\vR') 
=&
\left[ \hp_i \hp_j \right] \,
\vf^2 \Nf \; 4\pi T \sum_{\vare_m > 0 }  
\frac{\Delta^2_\hp}{\Omega^2} 
\frac{2}{\vf} \,
\frac{ e^{\displaystyle -\kappa_u |\Del s|} }{2\pi |\Del s| } 
\\
&\times \left[ \left(1-e^{-\kappa_u |s_<|} \right)^2 - \frac{\Omega^2}{\vare_m^2} e^{-2\kappa_u |s_<|} \right],
\end{split} \label{eq:K1} \end{align}
where $\vare_m = \pi T (2m+1)$ are the Matsubara energies, 
$\kappa_u = 2\Omega/\vf$ and $\Omega = \sqrt{\vare_m^2 + \Delta_\hp^2}$;
also $\Del s = s_\vR - s_{\vR'}$ is the trajectory distance between the two points,
and $s_< = \mbox{min}(y,y')/|\hp_y|$ is the trajectory coordinate of the point, $\vR$ or $\vR'$, closest to the surface.
For the reflection path ($\hp' = \ul\hp = \hp - 2\hat y (\hat y\cdot \hp)$)
\begin{align}\begin{split}
K^{\circled{2}}_{ij}(\vR,\vR') 
=
- \left[ \hp_i \ul\hp_j \right] \,
\vf^2 \Nf \; 4\pi T \sum_{\vare_m > 0 }  
\frac{\Delta^2_\hp}{\vare_m^2} 
\frac{2}{\vf} \,
\frac{ e^{\displaystyle -\kappa_u |\Del s | } }{2\pi |\Del s|}, 
\end{split} \label{eq:K2} \end{align}
where the overall minus sign is due to the 
fact that at the integration and observation points the order parameter 
has opposite signs $\Delta_{\ul\hp} = -\Delta_\hp$.  
This reflection involving the sign-change of the order parameter also leads 
to the zero-energy Andreev surface states.\cite{Hu:1994vo} 
The characteristic bound states term, proportional to $\Delta_\hp^2/\vare_m^2$, gives an overall $1/T$ temperature dependence of the kernel. 
The direct kernel in Eq.~(\ref{eq:K1}) may also show this $1/T$ dependence 
near the surface when the second term inside the square brackets dominates.

Pattern-forming instabilities are notorious for being technically challenging to analyze even at the level of linearised
equations \cite{PeschKramer}.
In what follows we 
work directly with the integral representation of the non-local physics. 
Since the unperturbed superconducting state is translationally invariant along the surface, we have
$\hat K(\vR,\vR')= \hat K(x_1-x_2, y_1, 0, y_2)$, and we may write the superflow free energy in terms of Fourier components of the phase, 
$ \chi(x,y) = C_{q_x} \chi(y) e^{+i q_x x} $, 
assuming the $\chi(y)$-profile to be real. We get
\begin{align}\begin{split}
\SF  = & \frac12 \int \frac{d q_x}{(2\pi)} |C_{q_x}|^2 \int\limits_0^\infty dy_1 \int\limits_0^\infty dy_2 \; \times 
\\
&\Big[ 
q_x^2  K_{xx} \; \chi(y_1) \chi(y_2) 
+  K_{yy} \; \chi'(y_1) \, \chi'(y_2) 
\\
& -i q_x K_{xy} \; \chi(y_1) \, \chi'(y_2) 
+ iq_x K_{yx} \; \chi'(y_1) \, \chi(y_2) 
\Big],
\end{split}\label{FEqx}\end{align}
where the prime denotes a derivative with respect to the $y$-coordinate. 
The kernel is a complicated function of several variables $K_{ij} = K_{ij}(q_x, y_1, y_2; T)$. To describe its most important features we 
use a center coordinate representation $y=(y_1+y_2)/2$, and integrate over the relative 
coordinate $\bar y = y_{1} - y_{2}$,
\be
\overline K_{ij}(q_x,y;T) = 
\int\limits_{-2 y}^{2 y } d \bar y \; K_{ij}\left(q_x,y+\frac12\bar y, y - \frac12 \bar y; T\right).
\label{Kavrg} \\
\ee
This averaged response is shown in Fig.~\ref{fig:S}{\bf b} as function of distance from the surface $y$, 
where we also include the $q_x$ multiplication factors to directly relate the kernel to the free energy. 
For $y \gtrsim L_y \approx 3\div 5 \xi_0$, the response is dominated by the direct path.
The off-diagonal components are zero and $K_{xx}$ and $K_{yy}$ are positive. 
Near the surface the diagonal components become negative, causing the instability, and large off-diagonal components appear.
All components have the $1/T$ low-temperature dependence near the surface. 
The sign-changing nature of $K_{ij}$, and its $T$-dependence, lead to fulfilment
of conditions (i) and (ii) for the phase crystal near the surface.
Moreover, exponential decay of the bound states into the bulk creates an asymmetric environment at the surface
with multiple $q_{0y}$ components contributing to the instability. Condition (iii) is thereby also satisfied.

\begin{figure*}[t]
\includegraphics[width=\linewidth]{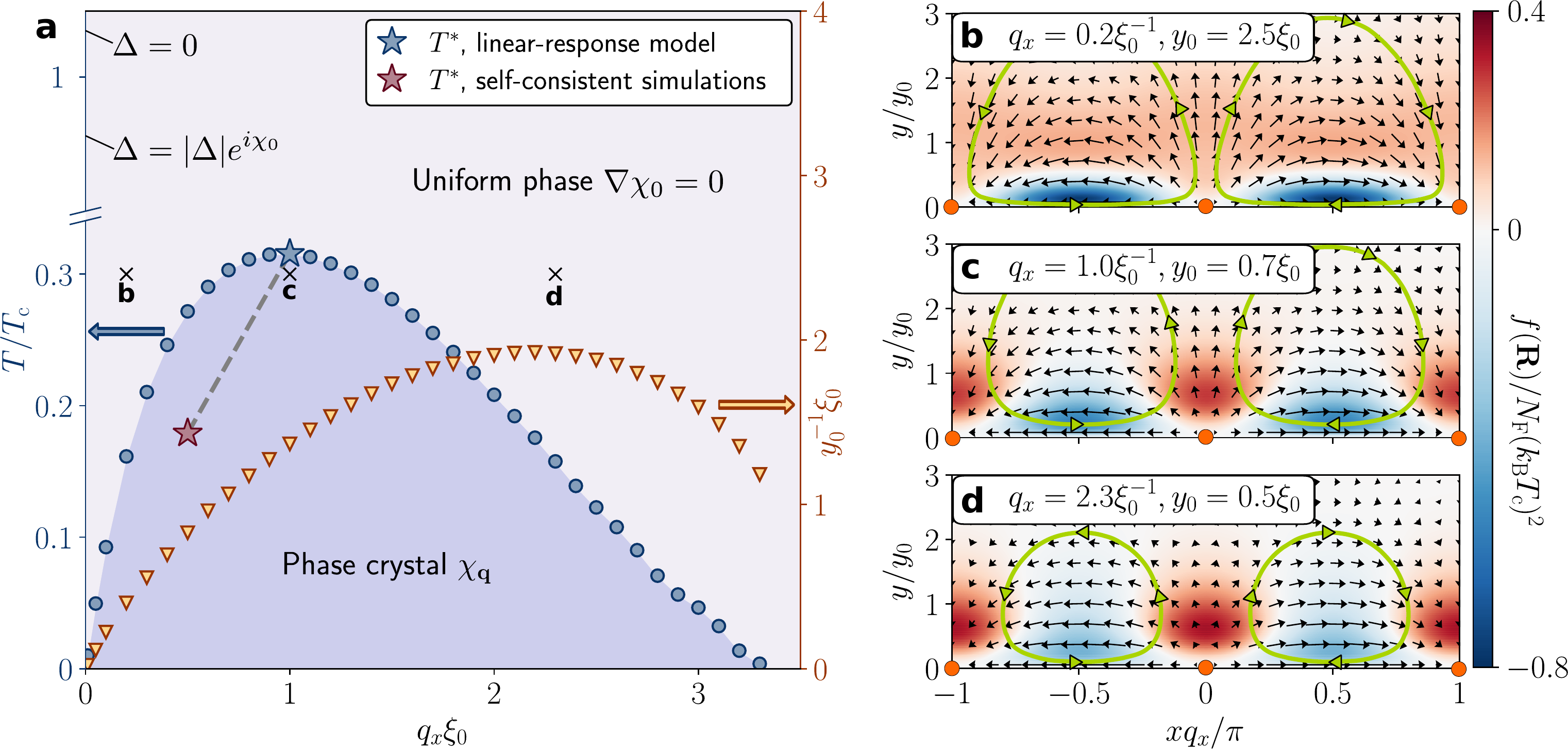}
\caption{
	{\bf a}, The $U(1)$ phase of the superconducting order parameter acquires periodic modulation below $T^*(q_x)$
	simultaneously breaking translational and time-reversal invariance of the $d$-wave superconducting state. 
	The highest-$T^*$ instability occurs at finite $q_x$, marked by the blue star. 
	The red star denotes the transition observed in a numerical self-consistent calculation \cite{Hakansson:2014uf}: 
	the lower $T^*$ is a result of the reduced spectral weight of zero-energy states due to order parameter suppression.
	In {\bf b}-{\bf d} we show the geometrical structure of the superflow $\vp_{\mathrm{s}}$ (black vector field) 
	and current streamlines (green loops)
	corresponding to physical solutions. 
	The background colors indicate
	distribution 
	of gradient energy gain and loss in the system. 
	At the optimal transition {\bf c} the overall energy is close to zero. 
	Increasing the pattern period, as in {\bf b}, leads to larger $y_0$ and deeper extension of currents into the bulk with
	bigger contributions from costly bulk gradient energies. 
	Making the pattern more compact, as in {\bf d}, increases the energy close to the surface.  
	In both {\bf b} and {\bf d} cases the loss in energy can only be compensated by lowering the temperature and
	thereby enhancing the negative bound states contribution through their $1/T$ dependence.
}
\label{fig:PD}
\end{figure*}

We perform a variational analysis of Eq.~(\ref{FEqx}) with an ansatz for the $y$-dependence of the phase 
decaying into the bulk on the scale of $y_0$,
\be
\chi(y) = \left(1+\frac{y}{y_0}\right) e^{-\frac{y}{y_0}} 
,\quad
\chi'(y) = -\frac{y}{y_0^2} e^{-\frac{y}{y_0}}.
\label{eq:chi_guess}
\ee
This choice is guided by considerations that there should be no currents deep in the sample, 
and we look for a state with no superflow in the $y$-direction at the surface. 
The latter condition is not a strict requirement, since the physical condition of no current across the boundary 
$j_y(y=0)=0$ is fulfilled automatically by the form of the total kernel $\hat K(\vR,\vR')$. 
This guess gives a good semi-quantitative result, but we note that to get the exact profile of $\chi(y)$ one has to perform a 
more sophisticated eigenvector analysis of the free energy Eq.~(\ref{FEqx}). 
%
For each wave vector $q_x$ and temperature $T$ we scan the variational parameter $y_0$ and find the minimum of 
the free energy. 
This minimum corresponds to the physical solution with currents satisfying $\dive \vj=0$.
The instability into the modulated-phase state with a non-zero $C_{q_x}$  occurs at a temperature where 
the minimum of $\SF$ crosses into negative values. 
The transition temperature $T^*(q_x)$ and the corresponding $y_0(q_x)$ are shown in Fig.~\ref{fig:PD}{\bf a}, for the $d$-wave case.
The highest transition temperature $T^* \sim 0.3 T_c$ occurs at finite modulation $q_x^* \approx \xi_0^{-1}$. 
By $x \to -x$ reflection symmetry there is degeneracy $(q_x,-q_x)$ that in the emerging state gives a real-valued phase and superflow 
\begin{align}\begin{split}
& \chi(x,y) \propto - \left(1+\frac{y}{y_0}\right) e^{-y/y_0} \, \cos q_x x \,,
\\
& \vp_s(x,y) 
\propto \left[ q_x \left(1+\frac{y}{y_0}\right) \sin q_x x, \;  \frac{y}{y_0^2} \cos q_x x \right] e^{-y/y_0}, 
\end{split}\end{align}
with the superflow exhibiting critical points $\vp_s=0$ at the surface,
as marked in Figs.~\ref{fig:PD}{\bf b}-{\bf d} by filled orange circles. 

\begin{figure*}[t]
\includegraphics[width=\linewidth]{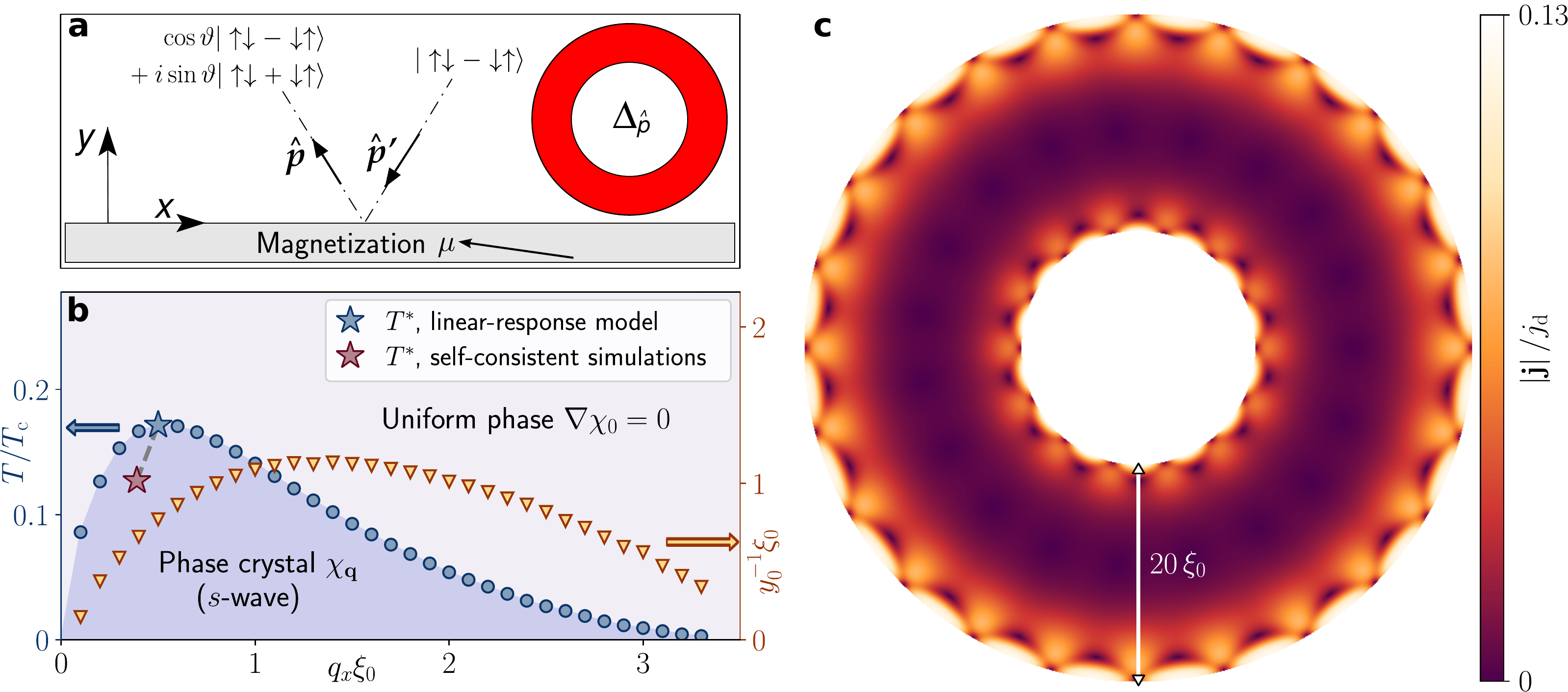}
\caption{
	{\bf a}, The phase crystallization can happen in conventional $s$-wave superconductors with magnetically-active surfaces
	that mix singlet and triplet correlations.\cite{EschrigABS}
	The zero-energy bound states are a result of spin mixing scattering processes with spin-mixing angle $\vartheta=\pi$. 
	{\bf b}, The general form of the surface superfluid kernel remains the same as in the $d$-wave case, and as a result the phase
	diagram looks similar. 
	{\bf c}, The fully self-consistent numerical result for the currents. For magnetic scattering the orientation of
	the surface is not important, and spontaneous currents can appear in any geometry. For the 2D annulus shown here, 
	the transition temperature is $T^*/T_c \approx 0.13 $. 
	Reduction of $T^*$ compared with the $d$-wave case is traced to angular dependence of the order parameter. 
}
\label{fig:Swave}
\end{figure*}

In the vicinity of the optimal transition, the instability temperature behaves as
\be
T^*(q_x) = T^* - \beta (q_x - q_x^*)^2 \,.
\ee
Such dependence is a characteristic ansatz in theories of weak crystallization\cite{Kats:1993}, 
where all the parameters are taken as phenomenological. 
We find $T^* \approx 0.3 T_c$, $q_x^* \approx 1.0/\xi_0 $ and $\beta \approx 0.15 T_c \xi_0^2$.
Here the appearance of a preferred finite phase modulation vector $q_x^*$ is the result of an interplay  
between terms in the free energy Eq.~(\ref{FEqx}) that in general have different 
dependence on the $y$-coordinates, $T$ and $q_x$. 
This physics can be crudely visualized by considering the superfluid free energy density, as shown in
Fig.~\ref{fig:PD}{\bf b}-{\bf d}.
\footnote{
The superfluid free energy density cannot be uniquely defined in non-uniform, and especially non-local,
systems. However, the two following definitions gave similar pictures:
$f_1(\vR) = \int d\vr \; \vp_s(\vR_+)^T \, \hat K(\vR_+, \vR_-) \, \vp_s(\vR_-)$ with $\vR_\pm = \vR \pm \vr/2$, 
and 
$f_2(\vR) = \vp_s(\vR) \cdot \vj(\vR) = \vp_s(\vR)^T \int d\vR' \, \hat K(\vR, \vR') \, \vp_s(\vR')$.
}
The key element is the dependence of the phase decay length $y_0$ on $q_x$, see Fig.~\ref{fig:PD}{\bf a} 
where we plot the inverse $y_0^{-1}(q_x)$. 
The superfluid response amplitudes grow with increasing $q_x$. 
At the same time, the peaks in $ q_x^2 K_{xx}$ and $q_x K_{xy,yx}$ move to smaller $y$, 
see Fig.~\ref{fig:S}{\bf b}.
This requires a smaller $y_0$ to control the current components to satisfy $\dive \vj=0$. 
Deviation of $q_x$ from its optimal value 
to smaller $q_x$, compare Fig.~\ref{fig:PD}{\bf b} with Fig.~\ref{fig:PD}{\bf c},
leads to a longer extent away from the surface of the phase oscillations which increases the bulk energy cost from $K_{xx}$ and $K_{yy}$.
On the other hand, a deviation to larger $q_x$ gives a small $y_0$ which results in a large cost
due to off-diagonal $K_{xy,yx}$ components, compare Fig.~\ref{fig:PD}{\bf d} with Fig.~\ref{fig:PD}{\bf c}.
The instability 
for non-optimal $q_x$ occurs at a lower temperature, where the $K_{xx}$-component becomes more negative
near the surface by virtue of its $1/T$ dependence,
which compensates for the energy increase in the other terms.   

From this analysis we may conclude that the non-local multi-component kernel leads to an intricate energy balance 
of the phase gradient terms in the free energy.
Because of the kernel structure, that fulfills the criteria (i)-(iii), a non-trivial phase crystallization
occurs at a particular $q_x^*\sim 1/\xi_0$. 
To this broad class of phase instabilities belong several previously described surface states with  
paramagnetic surface currents caused by spectral displacement of Andreev states.\cite{Fogelstrom:1997wp,Higashitani:1997bv}
That work assumed translational invariance of the superflow and currents along the surface, which guaranteed 
particle conservation $\dive \vj(\vR)=0$, 
but as a result required additional mechanisms of reducing superflow in the bulk. 
In semi-infinite systems one relies on the Meissner effect to screen the bulk superflow 
on the penetration depth length scale $\lambda$, which leads to 
$T^* \sim (\xi_0/\lambda) T_c$.\cite{Barash:2000vt,Lofwander:2000vh}
In slabs of width $D<\lambda$ the bulk contribution is obviously limited, resulting in spontaneous superflow below 
$T^* \sim (\xi_0/D) T_c$. \cite{VorontsovAB:2009ef}
In a similar fashion, we can interpret the phase crystal as self-screening of the loop currents over 
the surface region $L_y$ leading to $T^* \sim (\xi_0/L_y) T_c$.

A similar transition can appear in other anisotropic superconductors with reduced point group symmetry of the order parameter, 
such as polar $p$-wave which may also host a flat band of zero-energy surface fermions. 
Interestingly, phase crystallization can happen in conventional $s$-wave superconductors, 
where orbital pairbreaking scattering is absent. 
In this case, magnetically active interfaces can provide the proper environment for the phase instability, 
for example in superconductor-ferromagnetic structures. 
Such systems are being considered as important building blocks for spintronics applications, where  non-locality and
quantum coherence will play important roles.\cite{Eschrig:2011pt} 
As described in Appendix \ref{appC}, a similar form of the superfluid density tensor appear for 
$\vartheta=\pi$ spin mixing angle. The phase diagram and the result of a self-consistent calculation are shown in 
Fig.~\ref{fig:Swave}. 

The observable consequence of the spontaneous charge currents are magnetic fluxes near the surface.
The associated reconstruction of the edge ground state is important from another perspective, since it
can prevent realization of topological surface channels, as happens in topological insulators
\cite{Novelli:2019,Wang:2017}.
Moreover, softening of the surface superfluid density at some finite wavevector
can result in special features of surface transport, even without a fully developed instability.
This may be particularly relevant to transport in confined geometries.

Universal features of the pattern-formation phenomena in very different systems are manifested in 
the similarity of the phase diagram and the current patterns in Fig.~\ref{fig:PD} 
with those of the Rayleigh-B\'enard convection instability, 
which is also a result of geometrical constraints and conservation laws. 
There, the control parameter, instead of $T$, 
is the inverse Rayleigh ratio of buoyancy force to dissipative forces.\cite{CrossHohenberg}
We note that the 
convection roll currents 
in that case is due to an instability in a non-equilibrium driven system,
while the phase crystal is a second-order phase transition into a new ground state.

\section{Conclusions}

We have described a superconducting state where the global $U(1)$ phase
spontaneously forms a modulation in space, breaking continuous translational invariance.
The phase modulation results in a pattern of loop-currents and breaking of time-reversal symmetry. 
We have identified the general criteria (i)-(iii) that have to be met in order to get a
non-local superfluid density tensor
that favors phase crystallization.
Using microscopic theory, we showed that the circulating currents can appear at pair breaking surfaces of $d$-wave
superconductors.
In that case, quasiparticle reflections off the surface play a double role:
(a) they lead to a flat band of zero-energy Andreev bound states controlling signs of the superfluid components; 
and 
(b) they connect the $y$ and $x$ degrees of freedom at the level of the superfluid response 
resulting in preferred finite $q_x$-modulation of the superflow.
From previous numerical studies we know that this
state remains stable in external magnetic fields \cite{Holmvall:2018fl} and survives significant reduction of
spectral weight of bound states \cite{Holmvall:2019}.
Thus, one should expect that similar phenomena will arise in other condensates with zero-energy surface states.
To demonstrate this, we have stabilized the phase crystal in a conventional $s$-wave superconductor in contact with a
magnetically-active material, as can happen in hybrid superconductor-ferromagnet devices.
One particularly interesting scenario, for the future, would be to generate this phase in a bulk system.
The phase crystal presents an alternative vision of `supersolids' where phase-coherent states also spontaneously break 
translational symmetry, only in the amplitude of the order
parameter.\cite{supersolidRMP,SupersolidGas,Botticher:2019,Chomaz:2019} 
More generally, our results indicate that non-local effects in broken-symmetry states, 
especially with multi-component order parameters or competing orders, can lead to new states of matter. 
Such prospects are supported by early\cite{Pippard1953} and more recent\cite{Koyama2013} investigations  
of non-local physics in superconductors, 
as well as research into pattern formation due to long-range non-locality in biological systems.\cite{Tanaka2003,Bressloff2008,GarciaMorales2008}



\section {Acknowledgements}
The computations were performed on resources at Chalmers Centre for Computational Science and Engineering (C3SE)
provided by the Swedish National Infrastructure for Computing (SNIC). We thank the Swedish Research Council for
financial support. P.H. acknowledges Chalmersska forskningsfonden for travel support.










\appendix

\section{Superfluid density near a surface}
\label{appA}
\begin{figure}[t]
\centering\includegraphics[width=0.98\linewidth]{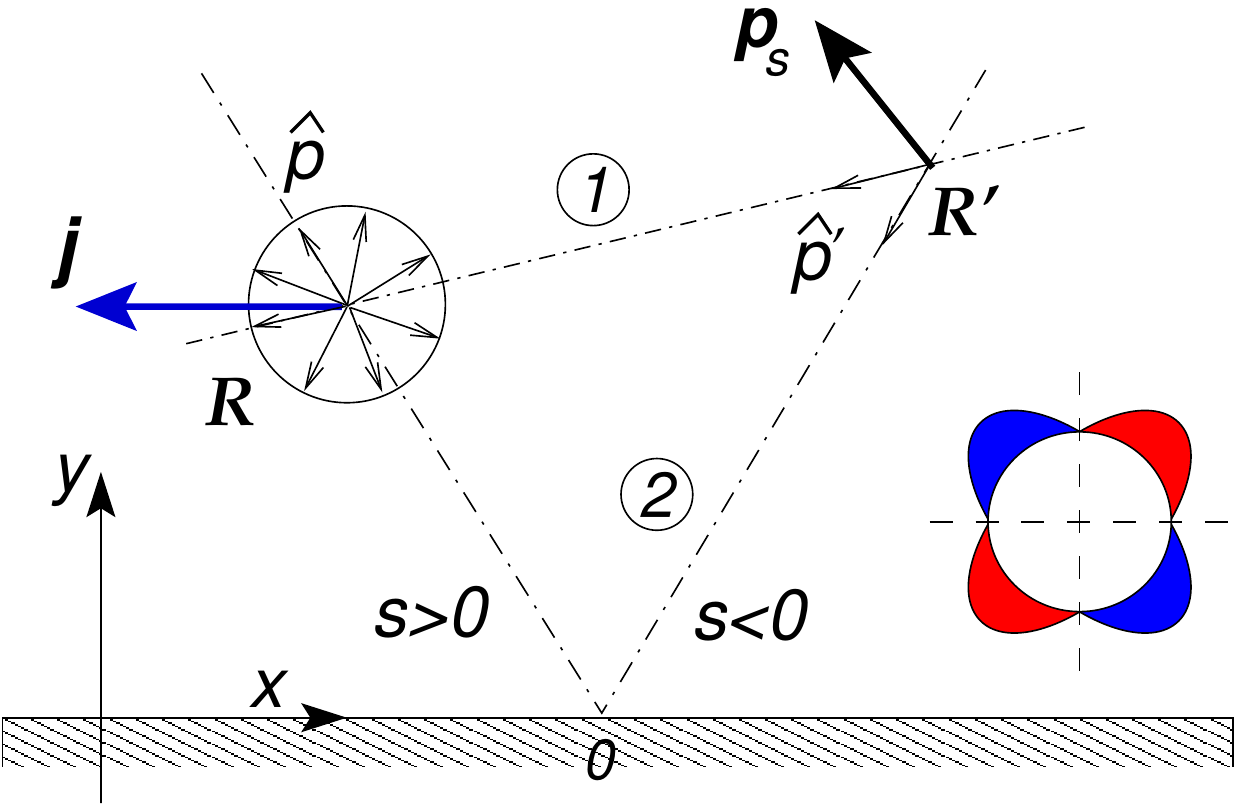}
\caption{
	The current at point $\vR$ is determined by quasiparticles carrying information about the superflow field $\vp_s$ 
	in the entire space. Near the surface, quasiparticles from point $\vR'$ can take two routes to get to point $\vR$:
	directly \circled{1}, and through a reflection off the interface \circled{2}. 
}
\label{fig:trajS}
\end{figure}

To find the superfluid response tensor we use a microscopic approach based on quasiclassical theory.\cite{Serene:1983vc}
Our starting point is the Eilenberger equation for the quasiclassical propagator $\hat g$
\be
[ (i\vare_m - \vv_f \cdot \vp_s) \hat \tau_3 - \hat\Delta(\vR,\vpf)\,,\, \hat g] 
+ i \vv_f \cdot \grad \hat g = 0 
\label{eqs:eil}
\ee
In this equation a spatially varying phase $\chi$ of the order parameter 
$\Delta = |\Delta| e^{i\chi(\vR)}$, was 
eliminated in favor of the superflow field $\vp_s = \frac12 \grad \chi$. 
This can always be done, if needed, by a gauge  transformation $\hat g \to  \hat U \hat g  \hat U^\dag$ 
with $\hat{U} = e^{i \hat \tau_3 \chi/2}$. 
The superflow is a function of position $\vp_s= \vp_s(\vR)$, and 
we consider a singlet mean-field order parameter $\Delta = \Delta(\vR,\vpf)$. 
The commutator-based Eilenberger equation is transformed into the Riccati-type equations for the coherence amplitudes\cite{Eschrig:2000ux}
\begin{align} \begin{split}
& i\vvf \cdot\grad \gamma + 2[i\vare_m - \vvf \cdot \vp_s] \gamma + \gamma \tilde\Delta \gamma + \Delta = 0,
\\
& i\vvf \cdot\grad \tilde\gamma - 2[i\vare_m - \vvf \cdot \vp_s] \tilde\gamma + \tilde\gamma \Delta \tilde\gamma + \tilde\Delta = 0.
\end{split} \label{eq:riccDE} \end{align}
These amplitudes conveniently parametrize the quasiclassical propagator,\cite{Schopohl1995,*Shelankov2000}
and are functions of position, momentum, and energy, $\gamma=\gamma(\vR, \vpf; \vare_m)$.
The two coherence amplitudes are related by symmetry, 
\be
\tilde\gamma(\vR, \vpf; \vare_m) = \gamma(\vR, -\vpf; \vare_m)^* \,,
\label{eq:symm_gamma}
\ee
that also applies to other tilde-related functions.
For the singlet real order parameter 
$\tilde\Delta(\vR,\vpf) \equiv \Delta^*(\vR,-\vpf) = \Delta(\vR,\vpf)$. 
We look at the current response due to a small but arbitrary superflow field $\vp_s=\vp_s(\vR)$, 
starting from a current-less background state $\Delta_0(\vR,\vpf)$ and the corresponding coherence amplitudes
$\gamma_0(\vR,\vpf;\vare_m)$.
The following linear response calculation is valid for any spatial profile of $\gamma_0(\vR,\vpf;\vare_m)$, and we
specify in the end its particular form.
The current at a point $\vR$ near the surface is calculated from the correction to the diagonal propagator 
$\delta g$, with $g = -i \pi \sgn(\vare_m) \frac{1 - \gamma \tilde\gamma}{1 + \gamma \tilde\gamma} $, as 
\be
\vj(\vR) 
= 2T \sum_{\vare_m > 0 } 2 \Nf \, \mbox{Re} \left\langle  \vf \hp \; \delta g(\vR, \vpf; \vare_m) \right\rangle_{\hp},
\ee
where  $\Nf$ is density of states at the Fermi level per spin projection, and 
$\left\langle \dots \right\rangle_{\hp} = \int d\hp /2\pi \, \dots $ denotes a cylindrical Fermi surface average, Fig.~\ref{fig:trajS}. 
In terms of linearised coherence amplitudes $\gamma = \gamma_0 + \gamma_1$ the propagator change due to small superflow is 
\be
{\delta g(\vR,\vpf; \vare_m)} 
=
{2 i \pi \sgn(\vare_m) }
\frac{\gamma_1 \tilde\gamma_0  + \gamma_0 \tilde\gamma_1 }{(1 + \gamma_0 \tilde\gamma_0)^2 }.
\label{eq:delta_g}
\ee
We first neglect the effect of the superflow on the amplitude of the order parameter, assuming that 
$\Delta(\vR) = \Delta_0(\vR)$ even in the current-carrying state, and linearise Eqs.~(\ref{eq:riccDE}) 
to find transport equations for the function $\gamma_1/(1+\gamma_0 \tilde\gamma_0)$,
\begin{align} \begin{split}
\hp\cdot\grad \frac{\gamma_1}{1 + \gamma_0 \tilde\gamma_0}  + \kappa 
\frac{\gamma_1}{1 + \gamma_0 \tilde\gamma_0}  = -2 i \; \hp \cdot \vp_s \; \frac{\gamma_0}{1 + \gamma_0 \tilde\gamma_0}.
\end{split} \label{eq:gamma_comb_transport} \end{align}
We get a similar equation for the tilde-analogue. 
The parameter 
\be
\kappa(\vR,\hp; \vare_m) \equiv  \frac{2}{\vf} \left[ 
\vare_m + \frac{ \gamma_0 \tilde\Delta_0 - \tilde\gamma_0 \Delta_0}{2i} \right] = \tilde \kappa
\,,\qquad 
\label{eq:kappadef}
\ee
determines the correlation length of the response. 
In a uniform state it reduces to $\kappa= 2\vf^{-1}\sqrt{\Delta_\hp^2 + \vare_m^2} \sim 1/\xi_0$. 

\begin{figure}[t]
\centering\includegraphics[width=0.9\linewidth]{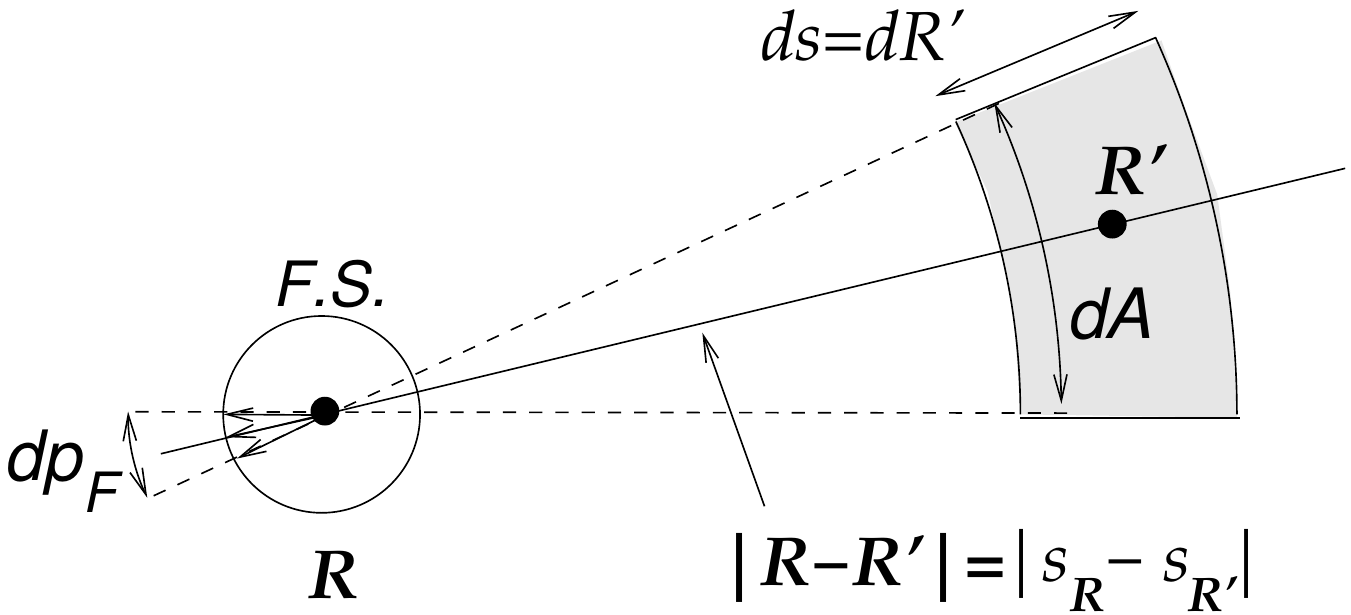}
\caption{
	The connection between spatial integral and the trajectory - Fermi surface integral. 
	A volume element $d^2R'$ in space 
	can be written in cylindrical coordinates as 
	$d^2R' = dA\, ds = |s_\vR-s_{\vR'}| d\vpf \, ds $, where $|s_{\vR}-s_{\vR'}|$ is the distance between
	points $\vR$ and $\vR'$ along a trajectory, $d\vpf$ is the angular integration over the Fermi surface. 
}
\label{fig:traject12}
\end{figure}

The solution of Eq. (\ref{eq:gamma_comb_transport}) along a quasiclassical trajectory $s$ is 
found, for positive $\vare_m$, by integration forward along the trajectory starting from zero
value in the bulk $\gamma_1(s=-\infty)=0$, where there is no superflow. We get
\begin{align}\begin{split}
\frac{\gamma_1}{1+\gamma_0 \tilde\gamma_0} (\vR, \hp; \vare_m) 
= - 2i  \int\limits_{-\infty}^{s_\vR} ds \; 
\exp\left(-\int_s^{s_\vR} \kappa(\rho) d\rho\right)  \; 
\\
\times \; \hp(s) \cdot\vp_s(\vR'(s)) \; 
\frac{\gamma_0 }{1+\gamma_0 \tilde\gamma_0} (s).
\end{split} \label{eqs:solution_gamma} \end{align}
To write the current at the observation point $\vR$ we need to integrate over all
trajectories coming into point $\vR$. 
By introducing a correlation function connecting two points, $\vR_{1}$ and $\vR_2$, by a quasiclassical trajectory 
$\hat\rho = (\vR_2-\vR_1)/|\vR_2-\vR_1|$, 
\be
C(\vR_2,\vR_1) =\frac{1}{2\pi |\vR_2-\vR_1|} \, \frac{2\vare_m}{\vf} \exp\left(-\int_{\vR_1}^{\vR_2} \kappa(\rho,\hat\rho) d\rho\right), 
\label{eq:C}
\ee
one can combine the Fermi surface average at the observation point and integration along trajectories into integration over 
all space $\vR'$, see Fig.~\ref{fig:traject12}, and write the current response as
\be
j_i(\vR) = \int d^2 R' \;  K_{ij}(\vR,\vR') p_{s,j}(\vR').
\label{eq:J=KP}
\ee

Inserting (\ref{eqs:solution_gamma}) into (\ref{eq:delta_g}) and using definition (\ref{eq:C}), 
the superfluid kernel is then given by 
\begin{align} \begin{split}
K_{ij}(\vR,\vR') 
= \vf^2 \Nf \; 8\pi T \sum_{\vare_m > 0 }  
\sum_{\circled{1},\circled{2}}  
\frac{1}{4\pi^2 \vare_m } \times 
\\
\times \mbox{Re} \left[ 
\hp_i \tilde f_0 (\vR,\hp)  C(\vR,\vR') f_0(\vR',\hp') \hp'_j 
+ \right. \\
+ 
\left. 
\hp'_j \tilde f_0(\vR',-\hp')  
C(\vR',\vR)
f_0 (\vR,-\hp) \hp_i 
\right],
\end{split} 
\label{eq:Kmicro}
\end{align}
where $f_0 $ and $\tilde f_0$ are off-diagonal propagators in the unperturbed state.  
In terms of coherence amplitudes $f_0 = -2 i \pi \, \sgn(\vare_m) \frac{\gamma_0}{1 + \gamma_0 \tilde\gamma_0} $. 
This kernel connects the observation point $\vR$ to the integration point $\vR'$. For each pair of points there are two paths,
one direct \circled{1} and one involving reflection at the surface \circled{2}, where we assumed mirror-like reflection, see Fig.~\ref{fig:trajS}.
The momentum direction $\hp$ at the observation point is given by the
trajectory direction $\vR' \to \vR$, and similarly for momentum at the integration point $\hp'$ (Fig.~\ref{fig:trajS}). 
These directions are different for the direct and reflected paths.

\section{Coherence amplitudes and propagators with a step-like order parameter} 
\label{appB}

\begin{figure}[t]
\centering\includegraphics[width=0.95\linewidth]{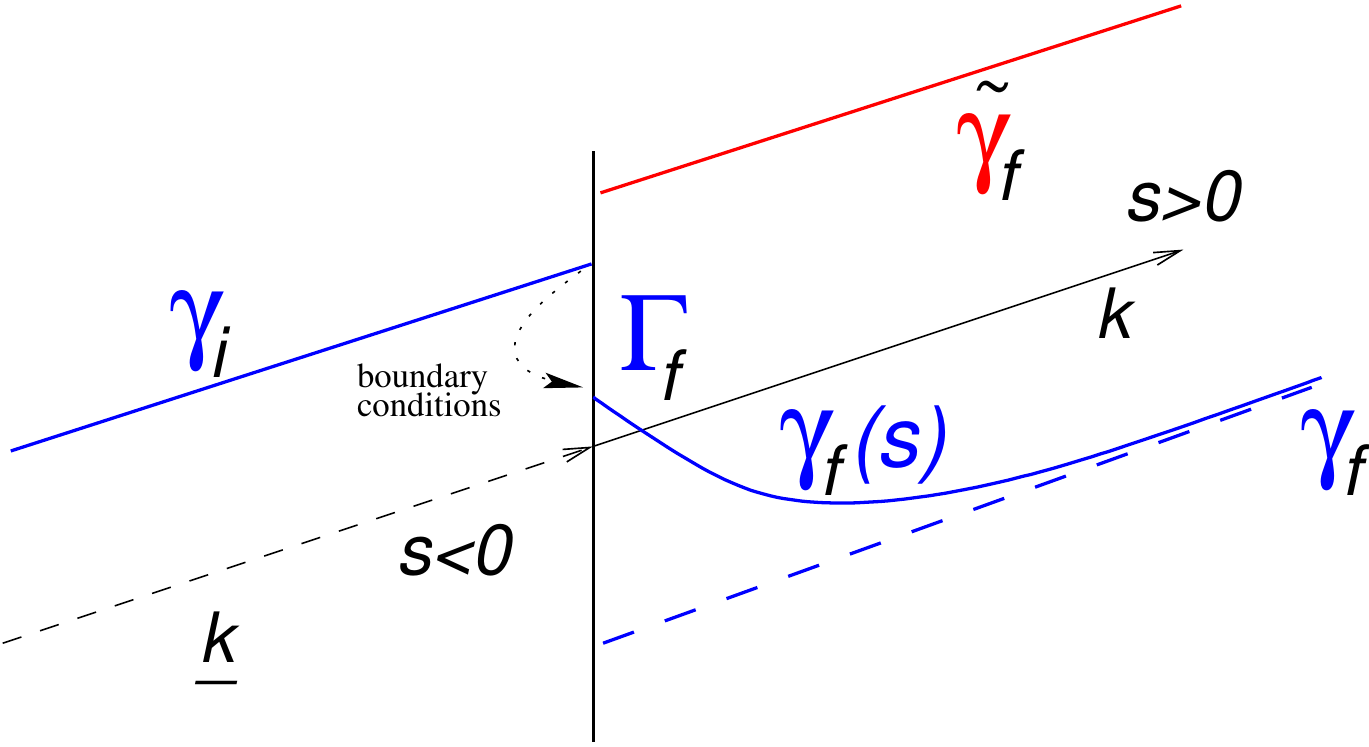}
\caption{
	The coherence amplitudes can be found analytically if we ignore suppression of the order parameter at the interface. 
	For each trajectory the order parameter sharply changes between $\Delta_i$ and $\Delta_f$ at
	$s=0$. In this case, $\gamma_i$ on incoming trajectory is a constant, 
	then a boundary condition 
	$\gamma_i \to \Gamma_f$ gives initial value that evolves to $\gamma_f$ on the outgoing part of trajectory.
	For typical non-magnetic specular scattering $\Gamma_f = \gamma_i$.
}
\label{fig:traject_gamma}
\end{figure}

Neglecting the suppression of the order parameter at the surface allows us to proceed further analytically.  
The bulk uniform coherence amplitude is 
\begin{align}\begin{split}
\gamma = i \frac{\Delta}{|\vare_m| + \sqrt{\Delta^2 + \vare_m^2}} \sgn(\vare_m) \,,
\\
\vare_m - i \gamma \Delta = \sgn(\vare_m) \sqrt{\vare_m^2 + \Delta^2} 
\end{split}\end{align}
Now consider, Fig.~\ref{fig:traject_gamma}, 
a (straightened) trajectory that for $s<0$ is in a region with the 
order parameter $\Delta_{\ul k} = \Delta_i$, 
and for $s>0$ is in the region with $\Delta_k = \Delta_f$ 
(e.g. for the most pairbreaking surface $\Delta_i = - \Delta_f$). 
Denote 
\begin{align}\begin{split}
\Omega_i = \sqrt{\Delta_i^2 + \vare_m^2} \;,\quad 
\kappa_{u,i} = \frac{2}{v_f}\sqrt{\Delta_i^2 + \vare_m^2} \,,
\\
\Omega_f = \sqrt{\Delta_f^2 + \vare_m^2} \;,\quad 
\kappa_{u,f} = \frac{2}{v_f}\sqrt{\Delta_f^2 + \vare_m^2} \,.
\end{split}\end{align}
Far away from the interface, the coherence amplitudes have their uniform bulk values (we assume $\vare_m > 0$, 
otherwise understand $\vare_m = |\vare_m|$ and add $\sgn(\vare_m)$ in front)
\begin{align}\begin{split}
\gamma_{i} = i \frac{\Delta_i}{\vare_m + \Omega_i}
\;,\qquad
\gamma_{f} = i \frac{\Delta_f}{\vare_m + \Omega_f}
\;,
\\
\tilde \gamma_{i} = -i \frac{\tilde\Delta_i}{\vare_m + \Omega_i}
\;,\qquad
\tilde \gamma_{f} = -i \frac{\tilde\Delta_f}{\vare_m + \Omega_f}
\end{split}\end{align}
For a sudden-step order parameter the amplitudes $\gamma_0,\tilde\gamma_0(s)$ can be found analytically, 
integrating Riccati equations (\ref{eq:riccDE}) in forward or backward direction, 
correspondingly. Including the sudden jump of the amplitudes at the surface according to the boundary condition, 
we get 
\begin{align}\begin{split}
\gamma_0(s<0) = \gamma_i \quad \rightarrow\quad 
\gamma_0(s=+0) = \Gamma_f \qquad\longrightarrow\qquad
\\
\gamma_0(s>0) = 
\gamma_{f} + \frac{ (1+\gamma_f \tilde\gamma_f) (\Gamma_f-\gamma_f) e^{-\kappa_{u,f} s}}{
	1+\gamma_f \tilde\gamma_f +(\Gamma_f-\gamma_f)\tilde\gamma_f (1-e^{-\kappa_{u,f} s})}
\end{split}\end{align}
and for tilde-function integrating backward:
\begin{align}\begin{split}
\tilde\gamma_0(s>0) = \tilde\gamma_f \quad\rightarrow\quad
\tilde\gamma_0(s=-0) = \tilde\Gamma_i \qquad\longrightarrow\qquad
\\
\tilde\gamma_0(s<0) = 
\tilde\gamma_{i} + \frac{ (1+\gamma_i \tilde\gamma_i) (\tilde\Gamma_i-\tilde\gamma_i) e^{\kappa_{u,i} s}}{
	1+\gamma_i \tilde\gamma_i +\gamma_i(\tilde\Gamma_i -\tilde\gamma_i) (1-e^{\kappa_{u,i} s})}
\end{split}\end{align}
The propagators on the trajectory are (e.g. for $s>0$)  
\begin{widetext}
\begin{align}\begin{split}
g_0 ( s>0) & = -i \pi \frac{1-\gamma_f(s) \tilde \gamma_f }{1+\gamma_f(s) \tilde \gamma_f }
= -i \pi \left[ 
\frac{1-\gamma_f \tilde \gamma_f }{1+\gamma_f \tilde \gamma_f }
\left(1-e^{-\kappa_{u,f} s} \right)
+ \frac{1-\Gamma_f \tilde \gamma_f }{1+\Gamma_f \tilde \gamma_f }
e^{-\kappa_{u,f} s} 
\right]
\end{split}\end{align}
and the off-diagonal component that enters the expression for the current response is 
\begin{align}\begin{split}
\frac{f_0(s>0)}{-2 i\pi} & =  \frac{\gamma_f(s)}{1+\gamma_f(s) \tilde\gamma_f}
= \frac{ \gamma_f }{1+\gamma_f \tilde \gamma_f } \left(1-e^{-\kappa_{u,f} s} \right)
  + \frac{\Gamma_f }{1+\Gamma_f \tilde \gamma_f } e^{-\kappa_{u,f} s} 
\\
\frac{\tilde f_0(s>0)}{2 i\pi} & =  \frac{\tilde \gamma_f}{1+\gamma_f(s) \tilde\gamma_f}
= \frac{ \tilde\gamma_f }{1+\gamma_f \tilde \gamma_f } \left(1-e^{-\kappa_{u,f} s} \right)
  + \frac{\tilde\gamma_f }{1+\Gamma_f \tilde \gamma_f } e^{-\kappa_{u,f} s} 
\\
& = \frac{ \tilde\gamma_f [ 1+\Gamma_f \tilde \gamma_f - (\Gamma_f- \gamma_f) \tilde \gamma_f e^{-\kappa_{u,f} s} ]}
{(1+\gamma_f \tilde \gamma_f)(1+\Gamma_f \tilde \gamma_f)  }  
\\
\frac{f_0(s<0)}{-2 i\pi} 
& = \frac{ \gamma_i }{1+\gamma_i \tilde \gamma_i } \left(1-e^{-\kappa_{u,i} |s|} \right)
  + \frac{\gamma_i }{1+\gamma_i \tilde \Gamma_i } e^{-\kappa_{u,i} |s|} 
\\
& = \frac{ \gamma_i [ 1+\gamma_i \tilde \Gamma_i - \gamma_i (\tilde\Gamma_i -\tilde\gamma_i)e^{-\kappa_{u,i} |s|} ] }
{(1+\gamma_i \tilde \gamma_i)(1+\gamma_i \tilde \Gamma_i)}  
\\
\frac{\tilde f_0(s<0)}{2 i\pi} 
  & = \frac{ \tilde\gamma_i }{1+\gamma_i \tilde \gamma_i } \left(1-e^{-\kappa_{u,i} |s|} \right)
  + \frac{\tilde\Gamma_i }{1+\gamma_i \tilde \Gamma_i } e^{-\kappa_{u,i} |s|} 
\end{split} \label{eqs:f} \end{align}
where we wrote the functions in several different ways, to cancel some terms later on. 

Notice the physical interpretation of the propagator form. For example, for $f_0(s>0)$ we have the same $\tilde\gamma_f$
in both terms since it is coming from $s=+\infty$, but the $\gamma$-amplitude can be either $\gamma_f$ far from the
reflection point or $\Gamma_f\leftarrow \gamma_i$ close to reflection points and they give rise to the two different terms in $f_0$. All
other expressions for $f$-functions follow the same pattern. 
The second term, that mixes $\Gamma_f$ and $\tilde\gamma_f$ in denominator, is
the one that mainly determines bound states effects. 
In both diagonal and off-diagonal items the continuum and the bound states contribution are nicely separated. 

\section{Current kernel without the order parameter suppression} 
\label{appC}

We use the results of Appendix \ref{appB} to calculate the current response kernel. 
First, we find $\kappa$ that determines the correlations extent in the current response: 
\begin{align}\begin{split}
\kappa(s) = \frac{2}{v_f} \left[ \vare_m + \frac{ \gamma_0 \tilde\Delta_0 - \tilde\gamma_0 \Delta_0}{2i} \right] 
=
\kappa_u \times \left\{ \begin{array}{l@{,\; \qquad}l} \displaystyle
1+ \frac{ (\Gamma_f-\gamma_f) \tilde\gamma_f e^{-\kappa_u s}}{
	1+\Gamma_f \tilde\gamma_f - (\Gamma_f-\gamma_f)\tilde\gamma_f e^{-\kappa_u s}}
	& s > 0 
\\ \displaystyle
1+ \frac{ \gamma_i(\tilde\Gamma_i- \tilde\gamma_i) e^{\kappa_u s}}{
	1+\gamma_i \tilde\Gamma_i - \gamma_i(\tilde\Gamma_i-\tilde\gamma_i) e^{\kappa_u s}} 
	& s < 0 
\end{array} \right.
\end{split}\label{eqs:kappa}\end{align}
Here we consider an order parameter orientation such that the amplitudes on the incoming and reflected parts of the
trajectory are the same, so $\kappa_{u,i} = \kappa_{u,f} = \kappa_{u}$. 
The generalization for different amplitudes can be easily carried out retaining indices
$\Omega_{i,f}$, $\kappa_{u;i,f}$ etc. 
This expression for $\kappa(s)$ is quite general and easy to integrate along trajectories, as required for correlation
functions $C(\vR,\vR')$ and $C(\vR',\vR)$. 
In both these functions integration goes from initial to final point as determined by the momentum direction, 
and is  
shown in Fig.~\ref{fig:C_integrals}. 

\begin{figure}[t]
\centering\includegraphics[width=0.7\linewidth]{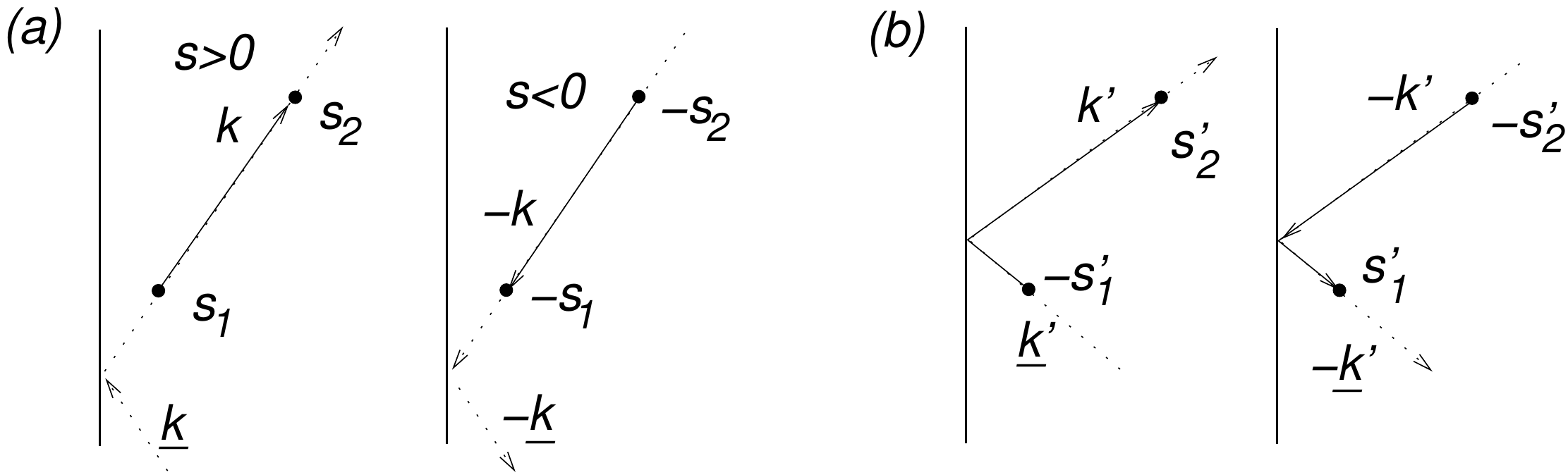}
\caption{
	The correlation functions that connect the integration point and the observation point along trajectories of type 
	\circled{1} (a) and type \circled{2} (b). 
}
\label{fig:C_integrals}
\end{figure}

For the case (a) both $s_1$ and $s_2$ are on the same side of the interface and $s_2$ is further away from the
interface than $s_1$, we have 
\be
s-out:\qquad 
C_{\circled{1}} \left(  \frac{1}{2\pi |s_2-s_1|} \frac{2\vare_m}{v_f} \right)^{-1} = 
\exp\left[-\int_{s_1}^{s_2} \kappa(\rho) d\rho \right] =  
\frac{1+\Gamma_f \tilde\gamma_f - (\Gamma_f-\gamma_f)\tilde\gamma_f e^{-\kappa_u s_1}} 
{1+\Gamma_f \tilde\gamma_f - (\Gamma_f-\gamma_f)\tilde\gamma_f e^{-\kappa_u s_2}} e^{-\kappa_u |s_2-s_1|} 
\label{eqs:C1a}
\ee
If we reverse the trajectory the signs of $s$ change (so that $s_1$ and $s_2$ determine absolute distance to the
surface)
\be
s-in:\qquad 
C_{\circled{1}} \left( \frac{1}{2\pi |s_2-s_1|} \frac{2\vare_m}{v_f} \right)^{-1} = 
\exp\left[-\int^{-s_1}_{-s_2} \kappa(\rho) d\rho \right] =  
\frac{1+\gamma_i \tilde\Gamma_i - \gamma_i(\tilde\Gamma_i-\tilde\gamma_i) e^{-\kappa_u s_1}} 
{1+\gamma_i \tilde\Gamma_i - \gamma_i(\tilde\Gamma_i-\tilde\gamma_i) e^{-\kappa_u s_2}}  e^{-\kappa_u |s_2-s_1|} 
\label{eqs:C1b}
\ee

For the (c) case we break the integral into two parts for in and out $s-in-out:$
\be
C_{\circled{2}} \left( \frac{1}{2\pi |s'_2+s'_1|} \frac{2\vare_m}{v_f} \right)^{-1} = 
\exp\left[-\int_{-s'_1}^{s'_2} \kappa(\rho) d\rho \right] =  
\frac{1+\gamma_f \tilde\gamma_f}{1+\Gamma_f \tilde\gamma_f - (\Gamma_f-\gamma_f)\tilde\gamma_f e^{-\kappa_u s'_2}} 
\frac{1+\gamma_i\tilde\gamma_i}{1+\gamma_i \tilde\Gamma_i - \gamma_i(\tilde\Gamma_i-\tilde\gamma_i) e^{-\kappa_u s'_1}}
e^{-\kappa_u (s'_2+s'_1)} 
\label{eqs:C2}
\ee
The denominators in (\ref{eqs:C1a}-\ref{eqs:C2}) 
will cancel numerators in some of the $f$-functions (\ref{eqs:f}) when combined in the kernel expression
(\ref{eq:Kmicro}). 
The numerators in (\ref{eqs:C1a}-\ref{eqs:C1b}) can be written as 
\begin{align}\begin{split}
1+\Gamma_f \tilde\gamma_f - (\Gamma_f-\gamma_f)\tilde\gamma_f e^{-\kappa_u s} = 
(1+\Gamma_f \tilde\gamma_f)(1-e^{-\kappa_u s} ) +(1+ \gamma_f \tilde\gamma_f) e^{-\kappa_u s} 
\\
1+\gamma_i \tilde\Gamma_i - \gamma_i(\tilde\Gamma_i-\tilde\gamma_i) e^{-\kappa_u |s|} 
= (1+\gamma_i \tilde\Gamma_i)(1- e^{-\kappa_u |s|}) + (1+\gamma_i \tilde\gamma_i) e^{-\kappa_u |s|} 
\end{split}\end{align}

For any given points $\vR$ and $\vR'$ we define two paths, direct and reflected, and each will have 
$\vR \to \vR'$ and $\vR' \to \vR$ contributions,
$\tilde f(\hp, \vR) C(\vR,\vR') f(\hp,\vR') + \tilde f(-\hp, \vR) C(\vR',\vR) f(-\hp,\vR)$. 
Let's denote by $\hk$ momentum away from the surface, and in this case we identify indices $f=\hk$, $i=\ul\hk$. 
The trajectory we are integrating $\gamma$-function goes from $s_1=s_<$ (point closest to the interface) to $s_2=s_>$
(point farthest from interface). 
For reverse trajectory we have $f=-\ul\hk$, $i=-\hk$ and integration happens from $-s_2$ to $-s_1$. 

The two terms give, after mentioned cancellations, for direct path
\begin{align}\begin{split}
& \frac{\tilde\gamma_0}{1 + \gamma_0 \tilde\gamma_0} C(\vR \leftarrow \vR') \frac{\gamma_0 }{1+\gamma_0 \tilde\gamma_0} 
+ 
\frac{\tilde \gamma_0 }{1+\gamma_0 \tilde\gamma_0} C(\vR'\leftarrow \vR) \frac{\gamma_0}{1+\gamma_0 \tilde\gamma_0} 
=
\frac{1}{2\pi |s_>-s_<|} \frac{2\vare_m}{v_f} \; \times \Bigg\{
\\
& \left[
\frac{\tilde\gamma_\hk }{1+\gamma_\hk \tilde\gamma_\hk} (1-e^{-\kappa_u s_<})
+ \frac{\tilde\gamma_\hk }{1+\Gamma_{\hk} \tilde\gamma_\hk} e^{-\kappa_u s_<}\right]
e^{-\kappa_u |s_>- s_<|}
\left[ \frac{\gamma_\hk }{1+\gamma_\hk \tilde\gamma_\hk} (1-e^{-\kappa_u s_<})
+ \frac{\Gamma_{\hk} }{1+\Gamma_{\hk} \tilde\gamma_\hk} e^{-\kappa_u s_<}\right] +
\\
& \left[
\frac{\tilde\gamma_{-\hk} }{1+\gamma_{-\hk} \tilde\gamma_{-\hk}} (1-e^{-\kappa_u s_<})
+ \frac{\tilde\Gamma_{-\hk} }{1+\gamma_{-\hk} \tilde\Gamma_{-\hk}} e^{-\kappa_u s_<}\right]
e^{-\kappa_u |s_>- s_<|}
\left[ \frac{\gamma_{-\hk} }{1+\gamma_{-\hk} \tilde\gamma_{-\hk}} (1-e^{-\kappa_u s_<})
+ \frac{\gamma_{-\hk} }{1+\gamma_{-\hk} \tilde\Gamma_{-\hk}} e^{-\kappa_u s_<}\right]
\Bigg\}
\end{split} \label{eq:Kdirect} \end{align}

For the reflected path this sum has a more compact form  that directly reflects the bound states factors 
\begin{align}\begin{split}
& \frac{\tilde\gamma_0}{1 + \gamma_0 \tilde\gamma_0} C(\vR \leftarrow \vR') \frac{\gamma_0 }{1+\gamma_0 \tilde\gamma_0} 
+ 
\frac{\tilde \gamma_0 }{1+\gamma_0 \tilde\gamma_0} C(\vR'\leftarrow \vR) \frac{\gamma_0}{1+\gamma_0 \tilde\gamma_0} 
=
\\
& 
=\frac{1}{2\pi |s'_>+s'_<|} \frac{2\vare_m}{v_f} \; \times \Bigg\{
\frac{ \tilde\gamma_{\hk'} \gamma_{\ul\hk'} }{( 1+ \Gamma_{\hk'}\tilde\gamma_{\hk'})( 1+ \gamma_{\ul\hk'}\tilde\Gamma_{\ul\hk'})} 
e^{-\kappa_u |s'_> + s'_<|}
+
\frac{\tilde\gamma_{-\ul\hk'} \gamma_{-\hk'} }{( 1+\gamma_{-\hk'} \tilde\Gamma_{-\hk'})( 1+\Gamma_{-\ul\hk'} \tilde\gamma_{-\ul\hk'})} 
e^{-\kappa_u |s'_> + s'_<|}
\Bigg\}
\end{split} \label{eq:Krefl} \end{align}
Note, that to generalize for inequivalent gap size on in-out trajectories we need to use appropriate 
$\kappa_u$ along given directions, e.g. $\kappa_u |s'_> + s'_<| \to \kappa_{u, \hk'} s'_> + \kappa_{u, \ul\hk'} s'_<$ for 
trajectory $\ul\hk' \to \hk'$ with reflection.  
These are completely general expressions for the one-component order parameters, where we neglect suppression of OP
amplitude near the surface, and assume specular scattering.  

\end{widetext}

We apply the developed formalism and approximations to a $d$-wave superconductor with maximally pairbreaking surface. 
In this case we have $\Delta_\hk = - \Delta_{\ul\hk}$ for all incident
trajectories, and $\gamma_{-\hk}=\gamma_{\hk} = -\gamma_{\ul\hk}$, 
$\tilde\gamma_{-\hk} =\tilde\gamma_{\hk} = \tilde\gamma_{\ul\hk}$, 
$\Gamma_\hk = \gamma_{\ul\hk} = - \gamma_\hk$, and two important combinations of the coherence
amplitudes are 
\be
\frac{1}{1+\gamma_\hk \tilde\gamma_\hk} = \frac{\vare_m + \Omega}{2\Omega}
\;,\qquad
\frac{1}{1+\Gamma_\hk \tilde\gamma_{\hk}} = \frac{\vare_m + \Omega}{2\vare_m}
\ee

The correlation coefficient Eq.~(\ref{eqs:kappa}) along a trajectory $s$ is 
\be
\kappa(s) 
=
\kappa_u 
\left[ \left( 1 -  e^{-\kappa_u |s|} \right) +  \frac{\Omega}{\vare_m} e^{- \kappa_u |s|} \right]^{-1},
\label{eq:kappa}
\ee
where $\kappa_u = 2\Omega/\vf$ and $\Omega = \sqrt{\vare_m^2 + \Delta_\hk^2}$.
The distance along a trajectory, measured from the surface, is $s = y /\hk_y$.
One uses these relations for coherence amplitudes in combinations (\ref{eq:Kdirect}) and (\ref{eq:Krefl})
to find the kernel (\ref{eq:Kmicro}) components, as given in the main text, for the direct path, 
Eq.~(\ref{eq:K1}),
and the reflection path,
Eq.~(\ref{eq:K2}), 
correspondingly.


%
Similar expressions for the superfluid density are valid for an $s$-wave superconductor with scattering 
at a specular magnetically-active surface. 
We use the  boundary conditions for coherence amplitudes\cite{Eschrig:2009ub} 
$$ \Gamma_{\hk} \, i\sigma_2 = {\cal M} \gamma_{\ul k} \, i\sigma_2 \tilde {\cal M} $$
with ${\cal M} = e^{i\vartheta \hat m \cdot {\bf \sigma} /2  }$
and $ \tilde {\cal M} = {\cal M}^*$.
Magnetic spin mixing leads to the bound states $\vare_b = \pm \Delta \cos (\vartheta/2)$, 
that result in zero energy states for 
$\vartheta = \pi$ and the boundary condition for coherence amplitudes 
$\Gamma_{\hk} = - \gamma_{\ul\hk}$.

\bibliography{fullPapersLib,NonLocalABS}

\end{document}